\documentclass[aps,twocolumn,prd,showpacs,nofootinbib]{revtex4}
\usepackage{amsmath}
\usepackage{graphicx}
\usepackage{dcolumn}
\usepackage{bm}
\usepackage{amssymb}
\usepackage{latexsym}
\begin{document}
\title[Thermodynamics of Abelian Gauge Fields in Hyperbolic Spaces]
{Thermodynamics of Abelian Gauge Fields in Real Hyperbolic Spaces}
\author{A. A. Bytsenko}
\address{Departamento de F\'{\i}sica, Universidade Estadual de Londrina, Caixa Postal 6001, Londrina-Parana, Brazil\,\, {\em E-mail address:} {\rm abyts@uel.br}}
\author{V. S. Mendes}
\address{Departamento de F\'{\i}sica, Universidade Estadual de Londrina, Caixa Postal 6001, Londrina-Parana, Brazil\,\, {\em E-mail address:} {\rm vsmendes@yahoo.com.br}}
\author{A. C. Tort}
\address{Departamento de F\'{\i}sica Te\'orica -- Instituto de F\'{\i}sica Universidade Federal do Rio de Janeiro, Caixa postal 68.528; CEP 21941-972, Rio de Janeiro, Brazil \,\, {\em E-mail address:} {\rm tort@if.ufrj.br}}
\date{November, 2003}
\thanks{}
\bigskip
\begin{abstract}
We work with $N-$dimensional compact real hyperbolic space 
$X_{\Gamma}$ with
universal covering $M$ and fundamental group $\Gamma$. Therefore, $M$
is the symmetric space $G/K$, where $G=SO_1(N,1)$ and $K=SO(N)$ is a
maximal compact subgroup of $G$. We regard $\Gamma$ as a discrete 
subgroup of $G$ acting isometrically on $M$, and we take $X_{\Gamma}$ to be the quotient space 
by that action: $X_{\Gamma}=\Gamma\backslash M = \Gamma\backslash G/K$.
The natural Riemannian structure on $M$ (therefore on $X$) induced by the
Killing form of $G$ gives rise to a connection $p-$form Laplacian 
${\frak L}_p$ on the quotient vector bundle (associated with an 
irreducible
representation of K). We study gauge theories based on abelian $p-$forms
on the real compact hyperbolic manifold $X_{\Gamma}$. 
The spectral zeta function 
related to the operator ${\frak L}_p$, considering only the co-exact 
part of 
the $p-$forms and corresponding to the physical degrees of freedom, 
can be 
represented by the inverse Mellin transform of the heat kernel. 
The explicit 
thermodynamic fuctions related to skew-symmetric tensor fields are 
obtained 
by using the zeta-function regularization and the trace tensor 
kernel formula 
(which includes the identity and hyperbolic orbital integrals). 
Thermodynamic quantities in the high and low temperature expansions are calculated and new entropy/energy ratios established. 
\end{abstract}
\pacs{04.70.Dy, 11.25.Mj}
\maketitle
\section{Introduction}
It is believed that the real universe is a form of
Friedman-Robertson-Walker models. However, the time-dependent metrics 
of these models brings 
about at least two inconvenient problems: (i) the definition of vacuum 
state in a time-dependent background; the background starts producing
particles in a continuous way, and therefore, vacuum states in the
Minkowskian sense do not exist; (ii) the all-important
thermodynamic investigation of the early universe becomes
senseless due to the lack of a proper definition of equilibrium
state \cite{Birrel&Davies}. Einstein static universes can meet the
inconveniences stated before. For these type of manifold, a vacuum
state can be unambiguously defined both locally and globally. It
can be shown that a closed universe has the same vacuum
energy and pressure as a static Einstein universe \cite{Ford1975}.
The radius of this universe is the instantaneous radius of the
Friedman-Robertson-Walker universe. Finite-temperature field theory in 
curved
backgrounds makes sense again. The thermodynamics of quantum
fields in an Einstein universe for some radius is equivalent to that
of an instantaneously static closed universe. 
The thermodynamics of positive curvature Einstein spaces was discussed
by several authors before. In particular the so-called entropy
bounds or entropy to thermal energy ratios were calculated and
compared with known bounds such as the Bekenstein bound or the
Cardy-Verlinde bound. For example, for a massless scalar field in
${\Bbb S}^3$ space this was done in \cite{Breviketal2003} and for a
massive scalar field in \cite{Elizalde&Tort2003}. Here we
wish to extend the evaluation of those type of bounds to the case
of skew-symmetric tensor fields in real compact hyperbolic spaces, a more 
delicate and involved task when compared with previous calculations in
${\Bbb S}^3$ space, as we shall see. This paper is divided as follows: In Section II 
we set the relevant mathematical tools of the quantum dynamics of exterior forms 
of real hyperbolic spaces; in Section III we apply Fried trace formula to the tensor 
kernel and derive the identity and the hyperbolic contributions; in Section IV 
the spectral functions associated with exterior forms are obtained; in Sections V and VI 
we establish the pertinent high and low temperature expansions and calculate the 
entropy bounds for the problem at hand. Comparison with results obtained for a massive 
scalar field in ${\Bbb S}^3$ and final remarks are left for the last Section. 
Throughout this paper we employ natural units, $\hbar =c=1$; 
Boltzmann constant $k_B$ is also set equal to the unity.  
\section{Quantum Dynamics of Exterior Forms of Real Hyperbolic Spaces}
Let $X_{\Gamma}$ be a $N-$dimensional real compact hyperbolic space 
with universal covering $M$ and fundamental group $\Gamma$.
Then we can represent $M$ as the symmetric space $G/K$, where 
$G=SO_1(N,1)$ 
and $K=SO(N)$ is a maximal compact subgroup of $G$. We regard 
$\Gamma$ as 
a discrete subgroup of $G$ acting isometrically on $M$, and we take 
$X_{\Gamma}$ to be the quotient space by that action: 
$X_{\Gamma}=\Gamma\backslash M= \Gamma\backslash G/K$.
Let $\tau$ be an irreducible representation of $K$ on a complex vector 
space
$V_\tau$, and form the induced homogeneous vector bundle 
$G\times_K V_\tau$
(the fiber product of $G$ with $V_\tau$ over $K$) over 
$M$. Restricting the $G$ action to $\Gamma$ we obtain the quotient 
bundle 
$E_\tau=\Gamma\backslash (G\times_KV_\tau)\rightarrow X_{\Gamma}$. 
The natural Riemannian structure on $M$ (therefore on $X_{\Gamma}$)
induced by the Killing form $(\;,\;)$ of $G$ gives rise to a connection
Laplacian $L$ on $E_\tau$. If $\Omega_K$ denotes the Casimir
operator of $K$ -- that is
$\Omega_K=-\sum y_j^2$,
for a basis $\{y_j\}$ of the Lie algebra ${\frak k}_0$ of $K$, where 
$(y_j\;,y_\ell)=-\delta_{j\ell}$, then $\tau(\Omega_K)=\lambda_\tau{}$ for a
suitable scalar $\lambda_\tau$. Moreover, for the Casimir operator $\Omega$
of $G$, with $\Omega$ operating on smooth sections $\Gamma^\infty E_\tau$ of
$E_\tau$ we have
\begin{equation}  \label{02}
L =\Omega-\lambda_\tau{}\,{\bf 1}\;;
\end{equation}
see Lemma 3.1 of Ref. \cite{Wallach}. For $\lambda\geq 0$ let
\begin{equation}  \label{03}
\Gamma^\infty\left(X\;,E_\tau\right)_\lambda= \left\{s\in\Gamma^\infty
E_\tau\left|- L s= \lambda s\right. \right\}
\end{equation}
be the space of eigensections of $L$ corresponding to $\lambda$.
Here we note that since $X_{\Gamma}$ is compact we can order the spectrum 
of $-L$ by taking $0=\lambda_0<\lambda_1<\lambda_2<\cdots$; $\lim_{j\rightarrow%
\infty}\lambda_j=\infty$. We shall focus on the 
case when $N=2k$ is even, and we shall specialize 
$\tau$ to be
the representation $\tau_p$ of $K=SO(2k)$ on 
$\Lambda^p {\Bbb C}^{2k}$,
say $p\neq k$. The case when $N$ is odd will be dealt with later. 
It will be
convenient moreover to work with the normalized Laplacian 
${\frak L} =-c(N)L$ where $c(N)=2(N-1)=2(2k-1)$. 
${\frak L}$ has spectrum 
$\left\{c(N)\lambda_j\;,m_j\right\}_{j=0}^\infty$ where the 
multiplicity $m_j$
of the eigenvalue $c(N)\lambda_j$ is given by
\begin{equation}  \label{04}
m_j={\rm dim}\;\Gamma^\infty\left(X\;,
E_{\tau_p}\right)_{\lambda_j}\;.
\end{equation}
Let $\omega_p,\, \varphi_p$ be exterior differential $p-$forms; then, 
the invariant inner product is defined by 
$(\omega_p, \varphi_p)\stackrel{def}{=}\int_{X_{\Gamma}}
\omega_p\wedge*\varphi_p$. The following
properties for operators and forms hold: $dd=\delta\delta=0$,\, $\delta
= (-1)^{Np+N+1}*d*$,\, **$\omega_p = (-1)^{p(N-p)}\omega_p$. 
The operators $d$ and $\delta$ are
adjoint to each other with respect to this inner product for
$p-$forms: $(\delta\omega_p, \varphi_p) = (\omega_p, d\varphi_p)$.
In quantum field theory the Lagrangian associated with $\omega_p$
takes the form: $d\omega_p\wedge *d\omega_p$ (gauge field)\,,\,
and $\delta\omega_p\wedge*\delta\omega_p$ (co--gauge field). The
Euler--Lagrange equations, supplied with the gauge, give: ${\frak
L}_p\omega_p =0\,,\,\,\delta\omega_p =0$ (Lorentz gauge);\,
${\frak L}_p\omega_p =0\,,\,\, d\omega_p =0$ (co--Lorentz gauge).
These Lagrangians provide a possible representation of tensor fields or
generalized abelian gauge fields. The two representations of
tensor fields are not completely independent, because of the
well--known duality property of exterior calculus which gives a
connection between star--conjugated gauge and co--gauge tensor
fields. The gauge $p-$forms are mapped into the co--gauge
$(N-p)-$forms under the action of the Hodge $*$ operator. The
vacuum--to--vacuum amplitude for the gauge $p-$form $\omega_p$
becomes \cite{Obukhov}:
\begin{eqnarray}
Z & = & {\cal N}\int D{\omega}\exp\left[
-(\omega, {\frak L}_p\omega)\right]
\nonumber \\
& \times &
\prod_{j=1}^p
\left({\rm Vol}_{p-j}({\rm det}{\frak L}_{p-j})^{(j+1)/2}
\right)^{(-1)^{j+1}}
\mbox{,}
\end{eqnarray}
where we need to factorize the divergent gauge group volume and
integrate over the classes of gauge transformations
($\omega \rightarrow \omega + d\phi$).
\section{The Trace Formula Applied to the Tensor Kernel}
The space of smooth sections $\Gamma^\infty E_\tau$ of
$E_\tau$ is just the space of smooth $p-$forms on $X$. We can
therefore apply the version of the trace formula developed by
Fried in \cite{Fried}. First we set up some additional
notation. For $\sigma_p$ the natural representation of $SO(2k-1)$
on $\Lambda^p {\Bbb C}^{2k-1}$, we have the corresponding
Harish--Chandra--Plancherel density given -- for a suitable
normalization of the Haar measure $dx$ on $G$ -- by
\begin{equation}  \label{07}
\mu_{\sigma_p(r)}= \frac{\pi}{2^{4k-4}[\Gamma(k)]^2} \left(
\begin{array}{c}
2k-1 \\
p
\end{array}
\right) P_{\sigma_p}(r)r \tanh(\pi r)\;,
\end{equation}
for $0\le p \le k-1$, where
\begin{eqnarray}  \label{08}
P_{\sigma_p}(r)& = & \prod_{\ell=2}^{p+1} \left[ r^2+\left(k-\ell+\frac{3}{2}
\right)^2 \right] \nonumber \\
& \times & \prod_{\ell=p+2}^{k} \left[ r^2+\left(k-\ell+\frac{1}{2}
\right)^2 \right]\;
\end{eqnarray}
is an even polynomial of degree $2k-2$. We have that $P_{\sigma_p}(r)=
P_{\sigma_{2k-1-p}}(r)$ and $\mu_{\sigma_p}(r)=\mu_{\sigma_{2k-1-p}}(r)$ for
$k\le p\le 2k-1$. Define the Miatello coefficients \cite{Miatello} $%
a_{2\ell}^{(p)}$ for $G=SO_1(2k+1, 1)$ by
\begin{equation}
P_{\sigma _{p}}(r)=\sum_{\ell =0}^{k-1}a_{2\ell }^{(p)}r^{2\ell }\;,\qquad
0\leq p\leq 2k-1\;.  \label{09}
\end{equation}
Let ${\rm Vol}(\Gamma \backslash G)$ will denote the integral of the
constant function ${\bf 1}$ on $\Gamma \backslash G$ with respect to the $G-$
invariant measure on $\Gamma \backslash G$ induced by $dx$.
For $0\leq p\leq N-1$ the Fried trace formula applied to kernel holds 
\cite{Fried}:
\begin{eqnarray}
{\rm Tr}\left( e^{-t{\frak L}_{p}}\right) & = & I_{\Gamma }^{(p)}({\cal K}_{t})
+I_{\Gamma }^{(p-1)}({\cal K}_{t}) 
\nonumber \\
& + & H_{\Gamma }^{(p)}({\cal K}_{t})
+H_{\Gamma }^{(p-1)}({\cal K}_{t})
\mbox{,}
\label{Fried}
\end{eqnarray}
where $I_{\Gamma }^{(p)}({\cal K}_{t}),\, H_{\Gamma }^{(p)}({\cal K}_{t})$ are the
identity and hyperbolic orbital integral respectively.
In the above formula
\begin{eqnarray}
I_{\Gamma }^{(p)}({\cal K}_{t}) & \overset{def}{=} & \frac{\chi (1){\rm Vol}
(\Gamma \backslash G)}{4\pi }\int_{{\Bbb R}}dr\mu _{\sigma
_{p}}(r) \nonumber \\
& \times &
e^{-t(r^{2}+b^{\left( p\right) }+\left( \rho _{0}-p\right) ^{2})},
\end{eqnarray}
\begin{eqnarray}
H_{\Gamma }^{(p)}({\cal K}_{t})& \overset{def}{=} & \frac{1}{\sqrt{4\pi t}}
\sum_{\gamma \in C_{\Gamma }-\{1\}}\frac{\chi (\gamma )}{j(\gamma )}
t_{\gamma }C(\gamma )\chi _{\sigma _{p}}(m_{\gamma }) \nonumber \\
& \times &
e^{-t\left( b^{\left(
p\right) }+\left( \rho _{0}-p\right) ^{2}\right) - t_{\gamma }^{2}/4t},
\end{eqnarray}
where $C_{\Gamma} \subset \Gamma$ is a complet set of representations 
in $\Gamma$ of its conjugacy classes, $C(\gamma)$ is a well 
defined function on $\Gamma - \{1\}$ (for more details see Ref. 
\cite{Williams}), $\rho_0=(N-1)/2$, $b^{(p)}$ are real constants, and 
$\chi_\sigma(m)={\rm trace} (\sigma(m))$ is the character $\sigma$
for $m\in SO(2k-1)$.

For $p\geq 1$ there is a measure $\mu_{\sigma}(r)$ corresponding
to a general irreducible representation $\sigma$.
Let $\sigma_p$ be the standard representation of $SO(N-1)$ on 
$\Lambda^p{\Bbb C}^{(N-1)}$. If $N=2k$ is even then
$\sigma_p\,\,(0\leq p\leq N-1)$ is always irreducible; if $N=2k+1$
then every $\sigma_p$ is irreducible except for $p=(N-1)/2=k$, in
which case $\sigma_k$ is the direct sum of two spin--$(1/2)$
representations
$\sigma^{\pm}:\,\,\sigma_k=\sigma^{+}\oplus\sigma^{-}$. For $p=k$ the
representation $\tau_k$ of $K=SO(2k)$ on $\Lambda^k {\Bbb C}^{2k}$
is not irreducible: $\tau_k=\tau_k^{+}\oplus\tau_k^{-}$ is the
direct sum of two spin--$(1/2)$ representations.
In the case
of the trivial representation ($p=0$, i.e. for smooth functions or
smooth vector bundle sections) the measure $\mu(r)\equiv
\mu_{0}(r)$ corresponds to the trivial representation. Therefore, 
we take $I_{\Gamma}^{(-1)}({\mathcal K}_t)
=H_{\Gamma}^{(-1)}({\mathcal K}_t)=0$. Since
$\sigma_0$ is the trivial representation, we have
$\chi_{\sigma_0}(m_{\gamma})=1$. In this case, formula (\ref{Fried})
reduces exactly to the trace formula for $p=0$
\cite{Wallach,Bytsenko1,Bytsenko2,Williams},
\begin{equation}
I_{\Gamma}^{(0)}({\mathcal K}_t)
=\frac{\chi(1)\mbox{vol}(\Gamma\backslash G)}
{4 \pi}\int_{\Bbb R}dr\,\mu_{\sigma_0}(r)e^{-t(r^2+b^{(0)}+\rho_0^2)}
\mbox{,}
\end{equation}
\begin{equation}
H_{\Gamma}^{(0)}({\mathcal K}_t)
=\frac{1}{\sqrt{4\pi t}}
\sum_{\gamma\in C_
\Gamma-\{1\}}\frac{\chi(\gamma)}{j(\gamma)}t_\gamma 
C(\gamma)e^{-t(b^{(0)}+\rho_0^2)- \frac{t_\gamma^2}{4t}}
\mbox{.}
\end{equation}
\section{The Spectral Functions of Exterior Forms}
The spectral zeta function related to the Laplace operator 
${\frak L}_{j}$ can be represented by
the inverse Mellin transform of the heat kernel 
${\cal K}_t={\rm Tr}\,\exp \left(-t{\frak L}_{j}\right)$.
Using the Fried formula, we can write the
zeta function as a sum of contributions:
\begin{eqnarray}
&&\zeta (s|{\frak L}_{j}) = \frac{1}{\Gamma (s)}
\int_{0}^{\infty }dt t^{s-1} 
\nonumber \\
&&\times \left( I_{\Gamma}^{(j)}({\cal K}_{t})+
I_{\Gamma }^{(j-1)}({\cal K}_{t})+H_{\Gamma }^{(j)}
({\cal K}_{t})+H_{\Gamma }^{(j-1)}({\cal K}_{t})\right)  
\nonumber \\
&&\equiv \zeta^{(N)}_{I}(s,j)+ \zeta^{(N)}_{I}(s,j-1)
\nonumber \\ 
&& + \zeta^{(N)}_{H}(s,j) + \zeta^{(N)}_{H}(s,j-1)
\mbox{.}
\end{eqnarray}
For the identity component we have
\begin{equation}
\zeta_{I}^{(N)}(s,j)=
\frac{V_{\Gamma}}{\Gamma (s)}
\int_{0}^{\infty }dt\,t^{s-1}\int_{{\Bbb R}}dr\,\mu_{\sigma _{j}}
e^{-t(r^{2}+\alpha_{j}^{2})},
\end{equation}
where 
$V_{\Gamma }=\chi (1){\rm Vol}\left( \Gamma \backslash G\right)/4\pi$, 
and we define
$
\alpha_{j}^{2}=b^{(j)}+\left( \rho_{0}-j\right)^{2}
$.
Replacing the Harish--Chandra--Plancherel measure, we obtain two
representations for $\zeta_{I}^{(N)}(s,j)$, which
holds for the cases of odd and even dimension. Thus,
\begin{eqnarray}
&& \zeta^{(2k)}_{I}(s,j) = \frac{V_{\Gamma}C_{2k}^{(j)}}
{\Gamma(s)}\sum_{\ell =0}^{k-1}a_{2\ell ,2k}^{(j)}
\nonumber \\
&& \times 
\int_{0}^{\infty }dt\,t^{s-1}\int_{\Bbb{R}}drr^{2\ell +1}
{ \rm tanh}(\pi r)\,e^{-t(r^{2}+\alpha _{j}^{2})}
\mbox{.}
\end{eqnarray}
Using the identities 
\begin{equation}
{\rm tanh}(\pi r) = 1- \frac{2}{1+e^{2\pi r}}
\mbox{,}
\end{equation}
\begin{equation}
\int_{0}^{\infty }\frac{drr^{2\ell -1}}{1+e^{2\pi r}}
=(-1)^{\ell -1}\frac{(1-2^{1-2\ell})B_{2\ell}}{4\ell}
\mbox{,}
\end{equation}
where $B_{\ell}$ is the $\ell-$th Bernoulli number, we get
\begin{eqnarray}
&& \zeta_{I}^{(2k)}(s,j) = \frac{V_{\Gamma }C_{2k}^{(j)}}
{\Gamma(s)}\sum_{\ell =0}^{k-1}a_{2\ell ,2k}^{(j)}   
\nonumber \\
&&\times \left[\Gamma (\ell +1)\Gamma (s-\ell -1)
\alpha_{j}^{-2s+2\ell +2}\right.   \nonumber \\
&&+\left. \sum_{n=0}^{\infty }\xi_{n\ell }\Gamma (s+n)
\alpha_{j}^{-2s-2n} \right]
\mbox{,}  
\label{zeta2}
\end{eqnarray}
where we have defined
\begin{equation}
\xi_{n\ell} \overset{def}{=}\frac{(-1)^{\ell +1}
\left(1-2^{-2\ell -2n-1}\right)}{n!(2\ell+2n+2)}B_{2\ell +2n+2}
\mbox{.}
\label{csi}
\end{equation}
In the odd dimensional case we get
\begin{eqnarray}
&& \zeta _{I}^{(2k+1)}(s,j) =  \frac{V_{\Gamma }C_{2k+1}^{(j)}}
{\Gamma (s)}\sum_{\ell =0}^{k}a_{2\ell ,2k+1}^{(j)}  
\nonumber \\
&&\times \int_{0}^{\infty}dt\,t^{s-1}
\int_{\Bbb{R}}drr^{2\ell }e^{-t(r^{2}+\alpha_{j}^{2})}
\nonumber \\
&& = \frac{V_{\Gamma }C_{2k+1}^{(j)}}
{\Gamma (s)}\sum_{\ell =0}^{k}
a_{2\ell, 2k+1}^{(j)}\Gamma \left( \ell+\frac{1}{2}\right) 
\nonumber \\
&&\times \Gamma \left( s-\ell -\frac{1}{2}\right) 
\alpha_{j}^{-2s+2\ell +1}
\mbox{.}
\end{eqnarray}
The hyperbolic component of the zeta function takes the form
\begin{equation}
\zeta _{H}^{(N)}(s,j) = \!\! \sum_{\gamma \in C_{\Gamma}-\{ 1\}}
\frac{\chi(\gamma) t_{\gamma}C( \gamma) 
\chi _{\sigma _{j}}(m_{\gamma})}{\sqrt{4\pi }\Gamma (s)
j(\gamma)}\int_{0}^{\infty }dt\frac{e^{-t\alpha _{j}^{2}-
\frac{t_{\gamma }^{2}}{4t}}}{t^{-s+ \frac{3}{2}}}
\mbox{.}
\end{equation}
Using the McDonald function,
\begin{equation}
K_{\nu}(z) =\frac{1}{2}\left( \frac{z}{2}\right)^{\nu }
\int_{0}^{\infty }dt\frac{e^{-t-\frac{z^{2}}{4t}}}
{t^{\nu+1}},  
\label{mac}
\end{equation}
where $|{\rm arg}\,z|<\pi /2$ and $\Re \,z^{2}>0$, we obtain

\begin{equation}
\zeta _{H}^{(N)}(s,j) = \!\! \sum_{\gamma \in C_{\Gamma}- \{ 1\}}
\frac{\chi (\gamma )t_{\gamma }^{2s}C(\gamma)
\chi _{\sigma_{j}}(m_{\gamma })}{\sqrt{\pi}\Gamma (s) j(\gamma )}
\frac{K_{-s+ \frac{1}{2}}(\alpha _{j}t_{\gamma })}
{(2\alpha t_{\gamma})^{s-1/2}}
\mbox{.}
\end{equation}
\section{The High Temperature Expansions}
Using the Mellin representation for the zeta function, we can
obtain useful formulas for the temperature dependent
part of the identity and hyperbolic orbital components of the 
free energy (see for detail Refs. \cite{Bytsenko1,Bytsenko2,Bytsenko3})
\begin{equation}
F^{(N)}_{I,H}(\beta,j) 
= -\frac{1}{2\pi i}\int\limits_{\Re z=c}\!\! \frac{dz}{\beta^z} 
\zeta (z) \Gamma(z-1)\zeta_{I,H} \left(\frac{z-1}{2},j\right)
\mbox{,}  
\label{re}
\end{equation}
where $\zeta (z)$ is the Riemann zeta function.
A tedious calculation gives the following results:
\begin{eqnarray}
F_{I}^{(2k)}(\beta,j ) &=&-\frac{V_{\Gamma }C_{2k}^{(j)}a_{2k-2,2k}^{(j)}}
{\sqrt{4\pi }}\Gamma (k)\zeta (2k+1)
\nonumber \\
&\times& \Gamma \left( k+\frac{1}{2}\right) \beta ^{-2k-1}
\nonumber \\
& - & \frac{V_{\Gamma }C_{2k}^{(j)}}{\sqrt{4\pi }}
\zeta (2k-1)\Gamma \left( k-\frac{1}{2}\right) 
\nonumber \\
& \times & \left[ a_{2k-4,2k}^{(j)}\Gamma (k-1)\right. 
\nonumber \\
& - & \left. a_{2k-2,2k}^{(j)}\Gamma \left( k \right) \right] 
\beta^{-2k+1}
\nonumber \\
& + & {\cal O}(\beta^{-2k+3}),
\end{eqnarray}
\begin{eqnarray}
F_{I}^{(2k+1)}(\beta,j ) & = & -\frac{V_{\Gamma}
C_{2k+1}^{(j)}a_{2k,2k+1}^{(j)}}{\sqrt{4\pi }}
\Gamma \left( k+\frac{1}{2}\right) 
\nonumber \\
& \times & \zeta (2k+2)\Gamma (k+1)\beta^{-2k-2}  
\nonumber \\
& - &\frac{V_{\Gamma }C_{2k+1}^{(j)}}{\sqrt{4\pi }}\zeta (2k)
\Gamma\left(k\right) 
\nonumber \\
& \times & \left[ a_{2k-2,2k+1}^{(j)}
\Gamma \left(k-\frac{1}{2}\right) \right.   
\nonumber \\
& - & \left. a_{2k,2k+1}^{(j)}\Gamma \left( k+\frac{1}{2}\right)
\alpha_{j}^{2}\right] \beta^{-2k}  
\nonumber \\
& + & {\cal O}\left( \beta^{-2k+2}\right) 
\mbox{.}
\end{eqnarray}
Note that the contribution associated to the hyperbolic orbital
component is negligible small. 
\subsection{The thermodynamic functions and the entropy bound}
In the context of the Hodge theory, the
physical degrees of freedom are represented by the co-exact part 
of the $p-$form. 
For $0\leq p\leq N-1$ the Fried trace formula \cite{Fried}
applied to the tensor kernel associated with co-exact forms 
has to be modified \cite{Bytsenko4,Bytsenko5,Bytsenko6}:
\begin{eqnarray}
&& {\rm Tr}\left( e^{-t{\frak L}^{(CE)}_{p}}\right)
= \sum_{j=1}^{p}\left(-1\right)^{j}
\left[ I_{\Gamma }^{(p-j)}({\cal K}_{t})+
I_{\Gamma }^{(p-1-j)}({\cal K}_{t})\right.   
\nonumber \\
&& +  \left. H_{\Gamma }^{(p-j)}({\cal K}_{t})+
H_{\Gamma }^{(p-1-j)}({\cal K}_{t})-b_{p-j}\right] 
\mbox{,}
\end{eqnarray}
where $b_{j}$ are the Betti numbers. Thus, the free energy becomes
\begin{eqnarray}
{\cal F}^{(N)}(\beta ) & = & \sum_{j=0}^{p}(-1)^{j}\left( F_{I}^{(N)}
(\beta,p-j) \right. 
\nonumber \\
& + & \left. F_{I}^{(N)}(\beta,p-j-1)\right) .
\end{eqnarray}
In the high temperature limit ($\beta \rightarrow 0$) we have
\begin{eqnarray}
{\cal F}^{(N)}(\beta ) &=&-A_{1}(N;\Gamma )\beta^{-N-1}  
\nonumber \\
& - & A_{2}(N;\Gamma )\beta^{-N+1}+{\cal O}(\beta^{-N+3}),
\end{eqnarray}
where for the even dimensional case,
\begin{eqnarray}
A_{1}\left( 2k;\Gamma \right) & = &\frac{V_{\Gamma }}{\sqrt{4\pi}}
\zeta \left(2k+1\right)\Gamma \left( k\right) 
\Gamma \left(k+\frac{1}{2}\right) \nonumber \\
& \times & 
C_{2k}^{(p)}a_{2k-2,2k}^{(p)}\,,  
\label{a}
\end{eqnarray}
\begin{eqnarray}
A_{2}(2k;\Gamma) & = & \frac{V_{\Gamma}}{\sqrt{4\pi}}
\zeta (2k-1) \Gamma (k-1)\Gamma \left(k-\frac{1}{2}\right)
\nonumber \\
& \times & 
C_{2k}^{(p)} \left( a_{2k-4,2k}^{(p)}
+ (k-1) a_{2k-2,2k}^{(p)} \right)
\mbox{,}
\label{b}
\end{eqnarray}
and for the odd dimensional case,
\begin{eqnarray}
A_{1}\left( 2k+1;\Gamma \right) &=&\frac{V_{\Gamma }}
{\sqrt{4\pi}}\zeta (2k+2)\Gamma \left(k+\frac{1}{2}\right) 
\Gamma (k+1)   
\nonumber \\
& \times & 
C_{2k+1}^{(p)}a_{2k,2k+1}^{(p)}
\mbox{,}
\label{c}
\end{eqnarray}
\begin{eqnarray}
A_{2}(2k+1;\Gamma ) = \frac{V_{\Gamma }}{\sqrt{4\pi }}
\zeta(2k)\Gamma \left(k-\frac{1}{2}\right)\Gamma(k)
\nonumber \\
\times 
C_{2k+1}^{(p)}\left(a_{2k-2,2k+1}^{(p)}
- \left(k-\frac{1}{2}\right)a_{2k,2k+1}^{(p)}
\alpha _{p}^{2}\right)
\mbox{.}
\label{d}
\end{eqnarray}
The entropy and the total energy can be obtained by means of the
following thermodynamic relations: $S^{(N)}(\beta )=\beta ^{2}{\partial}
{\cal F}^{(N)}(\beta )/\partial \beta $, $E^{(N)}(\beta )=
{\partial }(\beta {\cal F}^{(N)}(\beta))/{\partial }\beta$. 
Therefore,
\begin{eqnarray}
S^{(N)}\left( \beta \right) & = &\left( N+1\right) A_{1}
\left(N;\Gamma \right)
\beta^{-N}  \nonumber \\
&&+\left( N-1\right) A_{2}\left( N;\Gamma \right) \beta^{-N+2}
\nonumber \\
&&+{\cal O}\left( \beta^{-N+4}\right) ,
\end{eqnarray}
\begin{eqnarray}
E^{(N)}\left( \beta \right) & = &-NA_{1}\left( N;\Gamma \right) 
\beta^{-N-1}
\nonumber \\
&&-\left( N-2\right) A_{2}\left( N;\Gamma \right) \beta^{-N+1}
\nonumber \\
&&+{\cal O}\left( \beta^{-N+3}\right) ,
\end{eqnarray}
The ratio entropy/energy becomes
\begin{equation}
\frac{S^{(N)}(\beta)}{E^{(N)}(\beta)}=
\frac{N+1}{N}\beta +\frac{2}{N^{2}}
\frac{A_{2}(N;\Gamma )}{A_{1}(N;\Gamma )}\beta^{3}
+{\cal O}\left( \beta^{5}\right) 
\mbox{.}
\label{ratio}
\end{equation}
\section{The low temperature expansions}
In the low temperature limit we can use the following representation 
for the one-loop contribution to the free energy 
\cite{Bytsenko1,Bytsenko2}:
\begin{equation}
{\cal F}^{(N)}(\beta ) = -\frac{1}{\sqrt{\pi }}\sum_{n=1}^{\infty}
\int_{0}^{\infty }dt\,t^{-3/2}e^{-n^{2}\beta^{2}/4t}{\rm Tr}
e^{-t{\frak L}^{(CE)}_{p}}.
\end{equation}
For the identity contribution we have
\begin{eqnarray}
F_{I}^{(N)}(\beta,j ) & = &\frac{V_{\Gamma }}{\sqrt{\pi }}
\sum_{n=1}^{\infty }\int_{0}^{\infty }dt\, t^{-3/2}
\int_{{\Bbb R}}dr\mu _{{\sigma }_{j}}(r)   
\nonumber \\
&&\times e^{-t(\alpha _{j}^{2}+r^{2})-n^{2}\beta ^{2}/4t}
\mbox{.}
\end{eqnarray}

Therefore, the following formulas hold
\begin{eqnarray}
F_{I}^{(2k)}(\beta,j) & = & \frac{2V_{\Gamma }C_{N}^{(j)}}{\sqrt{\pi }}
\sum_{n=1}^{\infty }\sum_{\ell =0}^{k-1}a_{2\ell,2k}^{(j)}
\left[2^{3/2}\Gamma (\ell +1)\right. 
\nonumber \\
& \times & \alpha _{j}^{2\ell +3}
\frac{K_{\ell+\frac{3}{2}}(\alpha_{j}n\beta )}
{(\alpha _{j}n\beta )^{\ell +3/2}}
\nonumber \\
& - & (-1)^{\ell }\sum_{m=0}^{\infty }\frac{\left( 1-2^{-2\ell
-2m-1}\right) B_{2\ell +2m+2}}{m!\left( \ell +m+1\right)
2^{m-1/2}} 
\nonumber \\
& \times & \left. \alpha_{j}^{-2m+1}\frac{K_{-m+\frac{1}{2}}\left(
\alpha _{j}n\beta \right) }{\left( \alpha _{j}n\beta \right)^{-m+1/2}}
\right] ,  
\label{rep2}
\end{eqnarray}
\begin{eqnarray}
F_{I}^{(2k+1)}(\beta,j ) & = &\frac{2V_{\Gamma }C_{N}^{(j)}}{\sqrt{\pi }}
\sum_{n=1}^{\infty }\sum_{\ell =0}^{k}a_{2\ell ,2k+1}^{(j)}\Gamma
\left( \ell +\frac{1}{2}\right) 
\nonumber \\
&&\times (2\alpha _{j}^{2})^{\ell +1}\frac{K_{\ell +1}\left(
\alpha _{j}n\beta \right) }{\left( \alpha _{j}n\beta \right)^{\ell +1}}, 
\label{rep1}
\end{eqnarray}
\begin{eqnarray}
F_{H}^{(N)}(\beta,j ) & = & 2\alpha _{j}^{2}\sum_{n=1}^{\infty
}\sum_{\gamma \in C_{\Gamma }-\left\{ 1\right\} }\frac{\chi
\left( \gamma \right) t_{\gamma }C\left( \gamma \right) \chi
_{\sigma _{p}}\left(
m_{\gamma }\right) }{\pi j\left( \gamma \right) } 
\nonumber \\
&&\times \frac{K_{1}\left( \alpha _{N}\sqrt{n^{2}\beta
^{2}+t_{\gamma
}^{2}}\right) }{\left( \alpha _{N}\sqrt{n^{2}\beta ^{2}+t_{\gamma }^{2}}
\right) }.  
\label{rep3}
\end{eqnarray}
Using the asymptotic expansion for the McDonald function (\ref{mac}) 

\begin{eqnarray}
K_{\nu }\left( z\right)  &=&\sqrt{\frac{\pi }{2z}}e^{-z}\left[
\sum_{k=0}^{\ell -1}\frac{\Gamma \left( \nu +k+1/2\right)}
{(2z)^{k}k!\Gamma
\left( \nu -k-1/2\right) } \right.   
\nonumber \\
&&+\left. \frac{\theta \Gamma (\nu +\ell +1/2)}{(2z)^{\ell }\ell
!\Gamma (\nu -\ell +1/2)}\right] ,  
\label{kni}
\end{eqnarray}
for $\nu \in {\Bbb R}$, $z>0$, $\ell >\nu -1/2$ ($\ell =1,2,3,...$),  
$|\theta |\leq 1$ (see \cite{Gradshteyn}, page 963), in Eqs.
(\ref{rep2}), (\ref{rep1}) and (\ref{rep3}) we get the following result
for the entropy
\begin{eqnarray}
S^{(N)}\left( \beta \right) & = &\widetilde{A}_{1}\left( N;\Gamma \right)
\beta^{1/2}+\widetilde{A}_{2}\left( N;\Gamma \right) \beta^{-1/2}
\nonumber \\
&&+{\cal O}\left( \beta^{-3/2}\right) ,
\end{eqnarray}
where
\begin{eqnarray}
\widetilde {A}_{1}\left( 2k;\Gamma \right) & = & \sum_{\gamma \in
C_{\Gamma }-\left\{ 1\right\} }\frac{2\chi \left( \gamma \right)
t_{\gamma }C\left(\gamma \right) }{\sqrt{2\pi }j\left( \gamma \right)}
\chi_{\sigma_{p}}\left( m_{\gamma }\right)\alpha_{p}^{3/2}
\nonumber \\
& - & \frac{V_{\Gamma }}{2\pi}\sum_{\ell=0}^{k-1}\frac{\left(
-1\right)
^{\ell}\left( 1-2^{-2\ell-1}\right) B_{2\ell+2}}{\ell+1}
\nonumber \\
& \times &
C_{2k}^{(p)}a_{2\ell,2k}^{(p)}\alpha_{p}^{1/2}
\mbox{,}
\end{eqnarray}
\begin{eqnarray}
\widetilde{A}_{2}\left( 2k;\Gamma \right)  & = &
\frac{V_{\Gamma }}{\pi}C_{2k}^{(p)}a_{0,2k}^{(p)}\alpha_p^{3/2}
\nonumber \\
& + & \sum_{\gamma \in C_{\Gamma }-\left\{ 1\right\} }\frac{15\chi
\left( \gamma \right) t_{\gamma }C\left( \gamma \right)}
{8\sqrt{2\pi }J\left(\gamma \right)}
\chi_{\sigma_{p}}(m_{\gamma})\alpha_{p}^{1/2}
\nonumber \\
& + & \frac{3V_{\Gamma }}{4\pi }
\sum_{\ell=0}^{k-1}\frac{\left( -1\right) ^{\ell}\left(
1-2^{-2\ell-1}\right)
B_{2\ell+2}}{\ell+1}
\nonumber \\
& \times & C_{2k}^{(p)}a_{2\ell,2k}^{(p)}\alpha_{p}^{-1/2}
\mbox{,}
\end{eqnarray}
\begin{eqnarray}
\widetilde{A}_{1}\left( 2k+1;\Gamma \right)  &=&\sum_{\gamma \in
C_{\Gamma }-\left\{ 1\right\} }\frac{2\chi \left( \gamma \right)
t_{\gamma }C\left(
\gamma \right) }{\sqrt{2\pi }j\left( \gamma \right)}
\chi_{\sigma_{p}}(m_{\gamma })\alpha_{p}^{3/2}
\nonumber \\
& + & \frac{V_{\Gamma }}{\sqrt{2\pi }}
C_{2k+1}^{(p)}a_{0,2k+1}^{(p)}\alpha_{p}^{3/2},
\end{eqnarray}
\begin{eqnarray}
\widetilde{A}_{2}\left( 2k+1;\Gamma \right)  & = &\sum_{\gamma \in
C_{\Gamma }-\left\{ 1\right\} }\frac{\chi \left( \gamma \right)
t_{\gamma }C\left(
\gamma \right)}{\sqrt{2\pi }j\left( \gamma \right)}
\chi_{\sigma_{p}}(m_{\gamma })
\nonumber \\
& \times &
( 3\alpha_{p}^{1/2}+2\alpha_{p}^{3/2}) 
\nonumber \\
& + & \frac{15V_{\Gamma }}{8\sqrt{2\pi }}
C_{2k+1}^{(p)}a_{0,2k+1}^{(p)}\alpha_{p}^{1/2}
\mbox{.}
\end{eqnarray}
The energy is given by
\begin{eqnarray}
E^{(N)}(\beta)  & = &-\widetilde{A}_{1}\left( N;\Gamma \right)
\beta^{-1/2}-\widetilde{A}_{3}\left( N;\Gamma \right) \beta ^{-3/2}
\nonumber \\
& + & {\cal O}\left( \beta ^{-5/2}\right) ,
\end{eqnarray}
where
\begin{eqnarray}
\widetilde{A}_{3}\left( 2k;\Gamma \right)  &=&
-\frac{V_{\Gamma }}{\pi }
C_{2k}^{(p)}a_{0,2k}^{(p)}\alpha_{p}^{3/2}
\nonumber \\
& - & \frac{7}{4}\sum_{\gamma \in C_{\Gamma }-\left\{ 1\right\}
}\frac{\chi
\left( \gamma \right) t_{\gamma }C\left( \gamma \right) }{\sqrt{2\pi }
J\left( \gamma \right)}
\chi_{\sigma_{p}}(m_{\gamma })\alpha_{p}^{1/2}
\nonumber \\
& + & \frac{V_{\Gamma }}{4\pi }
\sum_{\ell=0}^{k-1}\frac{\left( -1\right)^{\ell}
\left(1-2^{-2\ell-1}\right)
B_{2\ell+2}}{\ell+1}
\nonumber \\
& \times &
C_{2k}^{(p)}a_{2\ell,2k}^{(p)}\alpha_{p}^{-1/2}
\mbox{,}
\end{eqnarray}
\begin{eqnarray}
\widetilde{A}_{3}\left( 2k+1;\Gamma \right)  & = &
\sum_{\gamma \in C_{\Gamma }-\left\{ 1\right\} }
\frac{\chi \left( \gamma \right) t_{\gamma }
C\left(\gamma \right) }{\sqrt{2\pi }j\left( \gamma \right) }
\chi_{\sigma_{p}}(m_{\gamma }) 
\nonumber \\
& \times &
(\alpha_{p}^{1/2}+2\alpha_{p}^{3/2})
\nonumber \\
& + &
\frac{7V_{\Gamma }}{8\sqrt{2\pi }}
C_{2k+1}^{(p)}a_{0,2k+1}^{(p)}\alpha_{p}^{1/2}
\mbox{.}
\end{eqnarray}
For the entropy/energy ratio in low temperature limit we get
\begin{eqnarray}
\frac{S^{(N)}(\beta)}{E^{(N)}(\beta)} & = &
\beta 
-\frac{\widetilde{A}_{3}(N;\Gamma )}
{\widetilde{A}_{2}(N;\Gamma)}
\left(\frac{\widetilde{A}_{2}(N;\Gamma )}
{\widetilde{A}_{1}(N;\Gamma )}
-\frac{\widetilde{A}_{3}(N;\Gamma }
{\widetilde{A}_{1}(N;\Gamma )}\right)\beta^{-1}
\nonumber \\
& + & {\cal O}(\beta^{-2})
\mbox{.}
\end{eqnarray}
\section{Conclusions}
We have considered gauge theories based on abelian $p-$forms in real compact hyperbolic manifolds. The explicit thermodynamic functions
associated with skew-symmetric tensor fields are obtained by
using zeta--function regularization and the trace tensor kernel
formula. Thermodynamic quantities in the low and high temperature limits 
were calculated.
We also have obtained the entropy/energy ratios (in both temperature limits). 
The dependence on the Miatello coefficients related to the structure of the
Harish--Chandra--Plancherel measure stems from the second term of the
expansion. 
In the case of scalar fields $(p=0)$ we have Eq. (\ref{ratio}) with
\begin{eqnarray}
\frac{A_{2}(2k;\Gamma )}{A_{1}(2k;\Gamma )} & = & 
\frac{2}{2k-1}\frac{\zeta (2k-1)}
{\zeta (2k+1)}
\nonumber \\
& \times &
\left( \frac{1}{k-1}\frac{a_{2k-4,2k}^{(0)}}{a_{2k-2,2k}^{(0)}
}-\alpha _{0}^{2}\right) ,
\end{eqnarray}
\begin{eqnarray}
\frac{A_{2}\left( 2k+1;\Gamma \right) }{A_{1}\left( 2k+1;\Gamma \right) }
& = &
\frac{1}{k}\frac{\zeta \left( 2k\right) }{\zeta \left( 2k+2\right)}
\nonumber \\
& \times &
\left( \frac{2}{2k-1}\frac{a_{2k-2,2k+1}^{\left( 0\right)
}}{a_{2k,2k+1}^{\left( 0\right) }}-\alpha _{0}^{2}\right) ,
\end{eqnarray}
where $\alpha _{0}^{2}=\rho _{0}^{2}+m^{2}$ 
($\alpha _{0}^{2}=\rho _{0}^{2}$ for the massless case).
For three-dimensional hyperbolic manifolds the Miatello coefficients 
read \cite{Bytsenko7}:
$a_{0}^{(0)}=a_{2}^{(0)}=1$ and therefore
\begin{equation}
\frac{S^{(3)}(\beta )}{E^{(3)}(\beta )} = \frac{4}{3}\beta + 
\frac{10}{3\pi ^{2}}
(2-\alpha _{0}^{2})\beta ^{3}+{\cal O}(\beta ^{5})
\mbox{.}
\end{equation}

This formula is in agreement with the result obtained in 
\cite{Elizalde&Tort2003}
where entropy bounds were calculated for spherical geometry
and where the dependence on geometry of the backgrounds also stems from 
the second term of the expansion.
\section*{Acknowledgements}
A.A.B. would like to thank Funda\c{c}\~ao de Amparo \`a Pesquisa do 
Estado de S\~ao Paulo (FAPESP/Brazil) and the Conselho Nacional de 
Desenvolvimento Cient\'{\i}fico e Tecnol\'ogico (CNPq/Brazil) for 
partial financial support, and the Instituto de F\'{\i}sica Te\'orica 
(IFT/UNESP) for kind hospitality. V.S.M. thanks CAPES for PhD grant. 

\end{document}